# Testing of Support Tools for Plagiarism Detection

**Version 1.0; 11th of February 2020**

Tomáš Foltýnek*, Mendel University in Brno, Czechia and University of Wuppertal, Germany, tomas.foltynek@mendelu.cz
Dita Dlabolová, Mendel University in Brno, Czechia, dita.dlabolova@mendelu.cz
Alla Anohina-Naumeca, Riga Technical University, Latvia, alla.anohina-naumeca@rtu.lv
Salim Razı, Canakkale Onsekiz Mart University, Turkey, salimrazi@comu.edu.tr
Július Kravjar, Slovak Centre for Scientific and Technical Information, julius.kravjar@cvtisr.sk
Laima Kamzola, Riga Technical University, Latvia, laima.kamzola@gmail.com
Jean Guerrero-Dib, Universidad de Monterrey, Mexico, jean.guerrero@udem.edu.mx
Özgür Çelik, Balikesir University, Turkey, ozgurcelik@balikesir.edu.tr
Debora Weber-Wulff, HTW Berlin, Germany, weberwu@htw-berlin.de

* Corresponding author

## Abstract

There is a general belief that software must be able to easily do things that humans find difficult. Since finding sources for plagiarism in a text is not an easy task, there is a wide-spread expectation that it must be simple for software to determine if a text is plagiarized or not. Software cannot determine plagiarism, but it can work as a support tool for identifying some text similarity that may constitute plagiarism. But how well do the various systems work? This paper reports on a collaborative test of 15 web-based text-matching systems that can be used when plagiarism is suspected. It was conducted by researchers from seven countries using test material in eight different languages, evaluating the effectiveness of the systems on single-source and multi-source documents. A usability examination was also performed. The sobering results show that although some systems can indeed help identify some plagiarized content, they clearly do not find all plagiarism and at times also identify non-plagiarized material as problematic.

## Declarations

### Availability of data and materials

Data and materials used in this project are publicly available from
http://www.academicintegrity.eu/wp/wg-testing/

## Competing interests

Several authors are involved in organization of the regular conferences *Plagiarism across Europe and Beyond*, which receive funding from Turnitin, Urkund, PlagScan and StrikePlagiarism.com. One team member received "Turnitin Global Innovation Awards" in 2015. These facts did not influence the research in any phase.

## Funding

This research did not receive any external funding. HTW Berlin provided funding for openly publishing the data and materials.

## Authors' contributions

TF managed the project and performed the overall coverage evaluation. DD communicated with the companies that are providing the systems. AAN and LK wrote the survey of related work. SR and ÖÇ wrote the discussion and conclusion. LK and DWW performed the usability evaluation. DWW designed the methodology, made all the phone calls, and improved the language of the final paper. All authors meticulously evaluated the similarity reports of the systems and contributed to the whole project. All authors read and approved the final manuscript. The contributions of others who are not authors are listed in the acknowledgements.

# Acknowledgments

We are deeply indebted to the contributions made to this investigation by the following persons:
- Gökhan Koyuncu and Nil Duman from the Canakkale Onsekiz Mart University (Turkey) uploaded many of the test documents to the various systems;
- Jan Mudra from Mendel University in Brno (Czechia) contributed to the usability testing, and performed the testing of the Czech language set;
- Caitlin Lim from the University of Konstanz (Germany) contributed to the literature review;
- Pavel Turčínek from Mendel University in Brno (Czechia) prepared the Czech language set;
- Esra Şimşek from the Canakkale Onsekiz Mart University (Turkey), helped in preparing the English language set;
- Maira Chiera from University of Calabria (Italy) prepared the Italian language set;
- Styliani Kleanthous Loizou from the University of Nicosia (Cyprus) contributed to the methodology design;
- We wish to especially thank the software companies that provided us access to their systems free of charge and patiently extended our access as the testing took much more time than originally anticipated.
- We also wish to thank the companies that sent us feedback on an earlier version of this report. We are not able to respond to every issue raised, but are grateful for them pointing out areas that were not clear.
2

# 1. Introduction

Teddi Fishman, former director of the International Centre for Academic Integrity, has proposed the following definition for plagiarism: "*Plagiarism occurs when someone uses words, ideas, or work products, attributable to another identifiable person or source, without attributing the work to the source from which it was obtained, in a situation in which there is a legitimate expectation of original authorship, in order to obtain some benefit, credit, or gain which need not be monetary*" (Fishman, 2009, p. 5). Plagiarism constitutes a severe form of academic misconduct. In research, plagiarism is included in the three "cardinal sins", FFP—Fabrication, falsification, and plagiarism. According to Bouter, Tijdink, Axelsen, Martinson, & ter Riet (2016), plagiarism is one of the most frequent forms of research misconduct.

Plagiarism constitutes a threat to the educational process because students may receive credit for someone else's work or complete courses without actually achieving the desired learning outcomes. Similar to the student situation, academics may be rewarded for work which is not their own. Plagiarism may also distort meta-studies, which make conclusions based on a number or percentage of papers that confirm or refute a certain phenomenon. If these papers are plagiarized, then the number of actual experiments is lower and conclusions of the meta-study may be incorrect.

There can also be other serious consequences for the plagiarist. The cases of politicians who had to resign in the aftermath of a publicly documented plagiarism case are well known, not only in Germany (Weber-Wulff, 2014) and Romania (Abbott, 2012), but also in other countries. Scandals involving such high-profile persons undermine citizens' confidence in democratic institutions and trust in academia (Tudoroiu, 2017). Thus, it is of great interest to academic institutions to invest the effort both in plagiarism prevention and in its detection.

Foltýnek, Meuschke, & Gipp (2019) identify three important concerns in addressing plagiarism:

1. **Similarity detection methods** that for a given suspicious document, are expected to identify possible source document(s) in a (large) repository;
2. **Text-matching systems** that maintain a database of potential sources, employ various detection methods, and provide an interface to users;
3. **Plagiarism policies** that are used for defining institutional rules and processes to prevent plagiarism or to handle cases that have been identified.

This paper focuses on the second concern. Users and policymakers expect what they call *plagiarism detection software*, but more exactly should be referred to as *text-matching software*, to use state-of-the-art similarity detection methods. The expected output is a report with all the passages that are identical or similar to other documents highlighted, together with links to and information about the potential sources. To determine how the source was changed and whether a particular case constitutes plagiarism or not, an evaluation by a human being is always needed, as there are many inconclusive or problematic results reported. The output of such a system is often used as evidence in a disciplinary procedure. Therefore, both the clarity of the report and the trustworthiness of its content are important for the efficiency and effectiveness of institutional processes.

There are dozens of such systems available on the market, both free and paid services. Some can be used online, while others need to be downloaded and used locally. Academics around the globe are naturally interested in the question: How far can these systems reach in detecting text similarities and to what extent are they successful? In this study, we will look at the state-of-the-art text-matching software with a focus on non-English languages and provide a comparison based on specific criteria by following a systematic methodology.

The investigation was conducted by nine members of the European Network for Academic Integrity (ENAI) in the working group TeSToP, **Te**sting of **S**upport **To**ols for **P**lagiarism Detection. There was no external funding available, the access to the various systems was provided to the research group free of charge by the companies marketing the support tools.

The paper is organized as follows. Section 2 provides a detailed survey of related work. Section 3 specifies the methodology used to carry out the research. Section 4 describes the systems used in the research. Section 5 reports the results acquired. Discussion and conclusion points are given at the end of the paper.

## 2. Survey of Related Work

Since the beginning of this century, considerable attention has been paid, not only to the problem of plagiarism, but also to text-matching software that is widely used to help find potentially plagiarized fragments in a text. There are plenty of scientific papers that postulate in their titles that they offer a classification, a comparative study, an overview, a review, a survey, or a comparison of text-matching software tools. There are, however, issues with many of the papers. Some, such as Badge and Scott (2009) and Marjanović, Tomašević, and Živković (2015), simply refer to comparative tests performed by other researchers with the aim of demonstrating the effectiveness of such tools. Such works could be useful for novices in the field who are not familiar with such automated aides, but they are meaningless for those who want to make an informed choice of a text-matching tool for specific needs.

Many research works offer only a primitive classification of text-matching software tools into several categories or classes. Others provide a simple comparative analysis that is based on functional features. They are usually built on a description of the tools as given on their official websites (e.g. Nahas, 2017; Pertile, Moreira, & Rosso, 2016), a categorization given in another source (e.g. Chowdhury & Bhattacharyya, 2016; Lukashenko, Graudina, & Grundspenkis, 2007; Urbina et al., 2010) or a study of corresponding literature and use of intelligent guess (e.g. Lancaster & Culwin, 2005). These types of research give a good insight into a broad scope of the functional features, focus, accessibility, and shortcomings of text-matching software. However, they are still incomplete for guiding the selection of a tool, as they do not evaluate and compare the performance of software systems and their usability from the viewpoint of end-users.

The most frequently mentioned categorizations are as follows:
- Software that checks text-based documents, source code, or both (Chowdhury & Bhattacharyya, 2016; Clough, 2000; Lancaster & Culwin, 2005; Lukashenko et al., 2007);
- Software that is free, private, or available by subscription (Chowdhury & Bhattacharyya, 2016; Lancaster & Culwin, 2005; Lukashenko et al., 2007; Nahas, 2017; Pertile et al., 2016; Shkodkina & Pacauskas, 2017; Urbina et al., 2010; Vandana, 2018);
- Software that is available online (web-based) or can be installed on a desktop computer (Lancaster & Culwin, 2005; Marjanović et al., 2015; Nahas, 2017; Pertile et al., 2016; Shkodkina & Pacauskas, 2017; Vandana, 2018);
- Software that operates intra-corpally, extra-corpally, or both (Lancaster & Culwin, 2005; Lukashenko et al., 2007; Marjanović et al., 2015).

Additionally, some researchers include unconventional comparative criteria. Pertile et al. (2016) indicate if a tool can make a citation analysis, a content analysis, structural analysis, or a paraphrase analysis. Lancaster and Culwin (2005) take into account the number of documents that are processed together to generate a numeric value of similarity and the computational complexity of the methods employed to find similarities. McKeever (2006) classifies text-matching software tools into search-based systems, systems performing linguistic analysis, software based on collusion detection, and systems for detecting software plagiarism.

Shkodkina and Pacauskas (2017) have defined 28 comparison criteria that are divided into four categories: affordability, material support, functionality, and showcasing. They compared three tools based on the criteria defined. However, it is not clear how the comparison was actually conducted, whether only by studying information available on product websites, or by trying out each tool. The descriptive part of their comparison does not contain references to information sources. Moreover, the set of criteria includes the ability of a tool to recognize different types of plagiarism (such as paraphrasing, translation, obfuscation, or self-plagiarism) and there are no indications of how these criteria were evaluated.

Shynkarenko and Kuropiatnyk (2017) have not compared available text-matching software tools, but they have defined more than 30 requirements for the development of such tools based on the analysis of other authors' works and the documentation of tools. They provide a comparison between the 27 tools mentioned in their paper and the defined requirements.

It is rather surprising that with the variety of research work on text-matching software, only a few of them address the performance and usability of these tools. Moreover, some of them do not make a comparative analysis of performance, but mainly check working principles and capabilities of the tools based on testing them on different kinds of submissions. At the end of the last century, Denhart (1999) published a lively discussion of his check of three systems.

He uploaded his senior thesis and a mini-essay made up of randomly selected sentences from four well-known authors with some slight paraphrasing to the systems. He found problems with properly quoted material and the inability to find many plagiarized sentences in the mini-essay. He also mentioned poor usability for one of the systems that otherwise had quite good performance results.

Culwin and Lancaster (2000) used a more carefully constructed text and checked four tools operating at that time using six sentences: four original sentences from two famous works widely available on the web, one paraphrased sentence from an essay available on a free essay site, and an original sentence from a newly indexed personal website. They checked the performance of tools and described if the text was found or not and at which sites. They also addressed some usability problems of systems for tutors and students.

Maurer, Kappe, and Zaka (2006) checked three tools in relation to verbatim plagiarism, paraphrasing, tabular information processing, translation plagiarism, image/multimedia processing, reference validity check, and a possibility to exclude/select sources. Despite that they are not describing the experiments in detail, there is evidence that they used a prepared sample of texts. These included a paragraph from proceedings that was paraphrased using a simple automatic word replacement tool, text compiled from documents available on the Internet, tabular information, and text in languages with special characters. They conclude that tools work reasonably well when plagiarized text is available on the internet or in other electronic sources. However, text-matching software fails to match paraphrasing plagiarism, plagiarism based on non-electronically available documents, and translation plagiarism. They also do not do well when processing tabular information and special characters.

Vani and Gupta (2016) used a small text fragment from the abstract of a scientific article and modified it based on four main types of obfuscation: verbatim plagiarism, random obfuscation, translation obfuscation, and summary obfuscation. Using the prepared text sample, they checked three tools and found that tools fail to find translation and summary obfuscations.

Křížková, Tomášková, and Gavalec (2016) made a comparative analysis of five systems completing two test series that used the same eight articles. The first test consisted of the articles without any modifications; the second test included manually modified articles by reordering words in the text. Their analysis consisted mainly of the percentage of plagiarism found and the time spent by the systems for checking the articles. They then applied multi-criteria decision-making for choosing the best system. However, there is no clear indication of the comparison goal, information about the already presented plagiarism in each of the articles, or how much plagiarism found by the systems matched the initially presented plagiarism. They also addressed usability by using a criterion "additional support" that includes a possibility to edit text directly on the website, multilingual checking, availability of vast information about plagiarism, etc.

Bull, Collins, Coughlin, and Sharp (2001) used a well-planned methodology and checked five systems identified through a survey of the academic staff from the higher education sector. They compared many functional features and also tested the performance of the tools. The criteria for evaluation contained among other issues the clarity of reports, the clarity of the instructions, the possibility to print the results, and the ease of interpreting the results that refer to the usability of tools. To test the performance they used eleven documents from six academic disciplines and grouped them into four categories according to the type of plagiarized material: essays from on-line essay banks, essays with verbatim plagiarism from the internet, essays written in collusion with others but with no internet material included, and essays written in collusion and containing some copied internet material. They tested the documents over a period of three months. In the end, they concluded that the tools were "*effective in identifying the types of plagiarism that they are designed to detect*" (Bull et al., 2001, p. 5). However, not all tools performed well in their experiments and they also reported on some anomalies in their results.

Chaudhuri (2008) examined only one particular tool and used 50 plagiarized papers from many different sources (freely available databases, subscription databases, open access journals, open sources, search engines, etc.) in different file formats. The researcher found that the tool is unable to match papers from subscribed databases, to process cited and quoted material, and articles from open access journals.

Luparenko (2014) tested 22 tools that were selected as popular ones based on an analysis of scientific literature and web sources. She considered many criteria related to functional specification (such as type, availability of free trial mode, need for mandatory registration at a website, number of users that have access to the program, database, acceptable file formats, etc.) and also checked the performance of the tools using one scientific paper in the Ukrainian language and another one in English. Moreover, the checking was done using three different methods: entering the text in the field of website, uploading a file, and submitting the URL of the article. She measured the checking time and evaluated the quality of the report provided by tools, as well as reported the percentage of unique text found in each of the articles.

The Croatian researchers Birkić, Celjak, Cundeković, and Rako (2016) tested four tools that are widely used in Europe and have the possibility to be used at the national and institutional level. They compared such criteria as the existence of an API (application programming interface) and the possibility to integrate it as a plug-in for learning management systems, database scope, size of the user community, and other criteria. The researchers tested the tools using two papers for each type of submission: journal articles, conference papers, master's and doctoral theses, and student papers. However, they did not include different types of plagiarism and evaluated the checking process with a focus on quote recognition, tool limitations, and interface intuitiveness.

Kakkonen and Mozgovoy (2010) tested eight systems using 84 test documents from several sources (internet, electronically unpublished books or author's own prepared texts, paper mills) that contained several types of plagiarism: verbatim copying, paraphrasing (e.g. adding more spaces, making intentional spelling errors, deleting or adding commas, replacing words by synonyms, etc.) and applying technical tricks. The technical tricks included the use of homoglyphs, which involve substituting similar-looking characters from different alphabets, and adding a character in a white-colored font as an empty space or including text as images. The authors provided a very detailed description of the experiments conducted and used a well-planned methodology. Their findings include problems with submissions from a paper mill, difficulties in identification of synonymous and paraphrased text, as well as finding the source for text obfuscated by technical tricks.

However, the most methodologically sound comparisons were conducted by Debora Weber-Wulff and her team (Weber-Wulff, Möller, Touras, & Zincke, 2013) between 2004 and 2013 (see http://plagiat.htw-berlin.de/start-en/). In their last testing experiment in 2013, the researchers compared 15 tools that were selected based on previous comparisons. The testing set contained both plagiarized and original documents in English, German, and Hebrew. The test set included various types of plagiarism from many different sources. They found serious problems with both false positives and false negatives, as well as usability problems such as many clicks needed for simple tasks, unclarity of reports, or language issues.

Summarizing the related work discussed in this chapter, it is worth mentioning that available studies on text-matching software:
- rarely address the evaluation of performance and usability of such tools, but mostly include a simple overview of their functional features, primitive categorization, or trivial comparisons;
- infrequently provide justification for the selection of tools based on well-defined reasons but often only mention the popularity of the tools;
- seldom use a well-planned scientific methodology and a well-considered corpus of texts in cases in which they evaluate the performance of the tools;
- do not report "*explicitly on experimental evaluations of the accuracy and false detection rates*" (Kakkonen & Mozgovoy, 2010, p. 139).

Taking into account the rapid changes in the field (some tools are already out of service, others have been continuously improved, and new tools are emerging) the need for comparative studies that in particular test the performance of the tools is constant. McKeever (2006, p. 159) also notes that "*with such a bewildering range of products and techniques available, there is a compelling need for up-to-date comparative research into their relative effectiveness*".

## 3. Methodology

The basic premise of this software test is that the actual usage of text-matching software in an educational setting is to be simulated. The following assumptions were made based on the academic experience of some members of the testing group before preparing for the test:

1. Students tend to plagiarize using documents found on the internet, especially Wikipedia.
2. Some students attempt to disguise their plagiarism.
3. Very few students use advanced techniques for disguising plagiarism (for example, homoglyphs).

4. Most plagiarizing students do not submit a complete plagiarism from one source, but use multiple sources.
5. Instructors generally have many documents to test at one time.
6. There are legal restrictions on instructors submitting student work.
7. In some situations the instructor only reviews the reports, submission is done either by the students themselves or by a teaching assistant.
8. Instructors do not have much time to spend on reviewing reports.
9. Reports must be stored in printed form in a student's permanent record if a sanction is levied.
10. Universities wish to know how expensive the use of a system will be on a yearly basis.

Not all of these assumptions were able to be put to test, for example, most systems would not tell us how much they charge, as this is negotiated on an individual basis with each institution.

## 3.1 Testing Documents

In order to test the systems, a large collection of intentionally plagiarized documents in eight different languages were prepared: Czech, English, German, Italian, Latvian, Slovak, Spanish, and Turkish. The documents used various sources (Wikipedia, online articles, open access papers, student theses available online) and various plagiarism techniques (copy & paste, synonym replacement, paraphrase, translation). Various disguising techniques (white characters, homoglyphs, text as image) were used in additional documents in Czech. The testing set also contained original documents to check for possible false positives and a large document to simulate a student thesis.

One of the vendors noted in pre-test discussions that they perceived Turnitin's exclusive access to publisher's databases as an unfair advantage for that system. As we share this view and did not want to distort the results, documents with restricted access were deliberately not included.

All testing documents were prepared by TeSToP team members or their collaborators. All of them were obliged to adhere to the written guidelines. As a result, each language set contained at least these documents:

- a Wikipedia article in a given language with 1/3 copy & paste, 1/3 with manual synonym replacement, and 1/3 manual paraphrase;
- 4–5 pages from any publicly available source in a given language with 1/3 copy & paste, 1/3 with manual synonym replacement, and 1/3 manual paraphrase;
- translation of the English Wikipedia article on plagiarism detection, half using Google Translate and half translated manually;
- an original document, i.e. a document which is not available online and has not been submitted previously to any text-matching software;

- a multi-source document in three variations, once as a complete copy & paste, once with manual synonym replacement and once as a manual paraphrase.

The basic multi-source document was created as a combination from five different documents following the pattern ABCDE ABCDE ABCDE ABCDE ABCDE, where each letter represents a chunk of text from a specific source. Each chunk was one paragraph (approx. 100 words) long. The documents were taken from Wikipedia, open access papers, and online articles. In some languages, there were additional documents included to test specific features of the systems.

Table 1 gives an overview of the testing documents and naming convention used for each language set.

**Table 1:** Testing documents and naming conventions

| Type | Name | | Description |
|---|---|---|---|
| Wikipedia | 01 | | Wikipedia article in the given language |
| | | a | Copy & paste |
| | | b | Manual synonym replacement |
| | | c | Manual paraphrase |
| Other sources | 02 | | 4–5 pages from a publicly available source in the given language |
| | | a | Copy & paste |
| | | b | Manual synonym replacement |
| | | c | Manual paraphrase |
| Translation from English | 03 | | English Wikipedia article on "Plagiarism detection" translated to the given language |
| Original document | 04 | | Original (previously unpublished online) document in the given language |
| Additional document(s) | 05, 06 | | Additional documents to check specific features (e.g. technical disguise, translation from a specific language, large files) |
| Multi-source document | 07 | | Five different documents in a given language combined |
| | | a | Copy & paste |
| | | b | Manual synonym replacement |
| | | c | Manual paraphrase |

Some language sets contained additional documents. Since many Slovak students study at Czech universities and the Czech and Slovak languages are very similar, a translation from Slovak to Czech was included in the Czech set and vice versa. There is also a significant Russian minority in Latvia so that a translation from Russian to Latvian was also included. The German set contained a large document with known plagiarism to test the usability of the systems, but it is not included in the coverage evaluation.

The documents were prepared in PDF, DOCX, and TXT format. By default, the PDF version was uploaded. If a system did not allow that format, DOCX was used. If DOCX was not supported, TXT was used. Some systems do not enable uploading documents at all, so the text was only copied and pasted from the TXT file. Permission to use the sources in this manner was obtained from all original authors. The permissions were either implicit (e.g. Creative Commons license), or explicit consent was obtained from the author.

## 3.2 Testing process

Between June and September 2018, we contacted 63 system vendors. Out of these, 20 agreed to participate in the testing. Three systems had to be excluded because they do not consider online sources and one because it has a word limit of 150 words for its web interface. In the next stage, the documents were submitted to the systems by authorized TeSToP members at a time unknown to the vendor. System default parameters were used at all times; if values such as minimum word run are discernable, they were recorded. After submission of the documents, one system withdrew from testing. Thus 15 systems were tested using documents in eight languages.

For evaluation, the following aspects were considered:

- coverage: How much of the known plagiarism was found? How did the system deal with the original text?;
- usability: How smooth was the testing process itself? How understandable are the reports? How expensive is the system? Other usability aspects.

To perform the coverage evaluation, the results were meticulously reviewed in both the online interface and the PDF reports, if available. Since the percentages of similarity reported do not include exact information on the actual extent of plagiarism and may even be misleading, a different evaluation metric was used. The coverage was evaluated by awarding 0–5 points for each test case for the amount of text similarity detected:

- 5 points: all or almost all text similarity detected;
- 4 points: a major portion;
- 3 points: more than half;
- 2 points: half or less;
- 1 point: a very minor portion;
- 0 points: one sentence or less.

For original work that produced false positives, the scale was reversed. Two or three team members independently examined each report for a specific language and discussed cases in which they did not agree. In some cases, it was difficult to assign points from the above-mentioned categories, especially for the systems which show only the matches found and hide the rest of the documents. If the difference between evaluators was not higher than 1 point, the average was taken. The interpretation of the above-mentioned scale was continuously discussed within the whole team.

To perform a usability evaluation, we designed a set of qualitative measures stemming from available literature (e.g. Badge & Scott, 2009; Chowdhury & Bhattacharyya, 2016; Hage Rademaker, & van Vugt, 2010; Martins, Fonte, Henriques, & da Cruz, 2014) and our experience. There were three major areas identified:

- Testing process;
- Test results;
- Other aspects.

Two independent team members assessed all systems in all criteria, only giving a point if the criteria are satisfactory and no points if not. After that, they discussed all differences together with a third team member in order to reach a consensus as far as possible. If an agreement was not possible, half a point was awarded. It a system offered such a functionality, but if the three researchers testing the systems were unable to find it without detailed guidance, 0.5 points were awarded.

Our testing took place between Nov 2018 and May 2019. During this time, we tested both coverage and usability. An additional test of multi-source document took place between August and November 2019. Since the present research did not benefit from any funding, the researchers were expected to fulfill their institutional workloads during the research period. Considering the size of the project team from various countries, we could make significant progress only during semester breaks, which explains the length of the testing process. It should be noted that we tested what the systems offered at the time of data collection. We used features that were allowed by the access given to us by the vendors.

The methodology was sent to all vendors, so that they were informed about the aim of the testing and other aspects of the process. The vendors were informed about categories of our testing (coverage criteria and usability criteria), as well as the fact we planned using documents in multiple languages.

Since the analysis and interpretation of the data are quite sensitive, we approached this period with the utmost care. As suggested by Guba and Lincoln (1989), member check is an effective technique for establishing the trustworthiness criteria in qualitative studies. Therefore, having analyzed and reported the data, we sent a preprint of the results to the vendors. Team members closely evaluated the issues raised by the vendors. Not all of them were able to be addressed in this paper, but as many as possible were incorporated. Because of the rigorous efforts to establish the validity of the results and the reliability of the study in this process, this study was further delayed.

# 4. Overview of Systems

In this chapter, a brief description of all of the web-based systems involved in the test is given. The information presented here is based on the information provided by the companies operating the systems—either from their website, or they were provided upon request by telephone or email using the list of questions documented in (Weber-Wulff, 2019). Links to the main pages of the systems can be found in the appendix.

The **Akademia** system presents itself as an anti-plagiarism system. It is intended for use at all levels of educational institutions and also for commercial institutions. The primary focus is on the region of Kosovo and Albania. The system was introduced in 2018. It is run by the company Sh.PK Academy Platform located in Pristina, Kosovo (Innovation Centre Kosovo, 2018).

**Copyscape** declares itself to be a plagiarism checker. The primary aim is to provide a tool for owners of websites to check if their original content was not used by others. They also provide a service of regular checks and email alerts. Copyscape, which started in 2004 (Greenspan, 2019), is operated by a private company, Indigo Stream Technologies Ltd., which is apparently based in Gibraltar. It does not have its own database but uses Google services to crawl the web.

**Docol©c** describes itself as a system for finding "*similarities between text documents on the Internet*" (Docol©c, 2019). It is intended for institutional use and focuses on German-speaking countries. According to the company, the system is used by more than 300 educational institutions in Austria, Germany, and Switzerland, plus around 20 universities worldwide and is integrated into the conference systems EDAS and OpenConf. Docol©c is operated by a private company, Docoloc UG (haftungsbeschränkt) & Co KG, based in Germany. It was developed in the years 2004-2005 at the University of Braunschweig— intended for personal use only. In 2006, it became available commercially. It uses MS Bing services to crawl the web and enables its customers to connect and browse their own databases. The license costs depend on the number of pages to be scanned per year and per institution.

**DPV** is part of the Slovenian National Portal of Open Science, which also provides an academic repository. The project, which is supported by the Slovenian higher education institutions, started in 2013. The detection software was developed by researchers from Slovenian universities. The operation of the system is partially funded by the European Union from the European Regional Development Fund and the Ministry of Education, Science and Sport (Ojsteršek et al., 2014).

**Dupli Checker** presents itself as a plagiarism checker. It is a free tool, but each search is limited to 1,000 characters). It does not focus on any specific users or purposes. There is no information about who operates it available at the website, we were also not able to receive such information when we asked directly via email. The website offers a variety of tools such as a paraphrasing tool and many search engine optimization (SEO) and other website management tools. Additionally, according to the statement found on their web site, they *"have delivered over 1,000,000 pages of high-quality content which attracts large amounts of traffic* *to [their] client's websites"* (Dupli Checker, 2019), so that it appears that they also offer a copywriting service.

The system **intihal.net** is operated by a Turkish private company, Asos Eğitim Bilişim Danışmanlik, and focuses on that language. According to direct information from the representatives, the system is being used by 50 Turkish universities and it has been operating approximately since 2017.

**PlagAware** is operated by a German private company, PlagAware Unternehmergesellschaft (haftungsbeschränkt). PlagAware states that it has 39,000 active users, and focuses on a wide range of customers—universities, schools and businesses, that are offered institutional licenses, and individuals, who can use purchased credits for individual checks of documents. They promise to perform the comparison with 10 billion online documents and reference texts provided by the user.

**Plagiarism Software** is operated by a private company settled in Pakistan. They focus on any type of individual users and claim to have 500,000 users. According to the information from the representative of the company, they started approximately in 2014 (on the web they claim to have seven years of experience which would date back to 2012) and they are using search engine services to browse the web. They offer five levels of pricing that differ according to the amount of the content being compared.

**PlagiarismCheck.org** presents itself as a plagiarism checking tool. It is operated by a company based in the United Kingdom and it has been on the market since 2011. Since around 2017 they are focusing on the B2B market. They state that they have more than 77,000 users in 72 countries. They use MS Bing for online searches and for the English language. The representatives claim they are able to do synonym detection. They provide three levels of institutional licenses.

**PlagScan** presents itself as a plagiarism checker. It is operated by the German company PlagScan GmbH and was launched in 2009. They state that they have more than 1,500 organizations as customers. Although they focus on higher education, high schools, and businesses, PlagScan is also available for single users. They search the internet using MS Bing, published academic articles, their so-called "Plagiarism Prevention Pool", and optionally a customer's own database. PlagScan offers multiple pricing plans for each type of customer, there are apparently also now options for a free trial.

**StrikePlagiarism.com** presents itself as a plagiarism detection system, operated by the Polish company Plagiat.pl. It provides its services to over 500 universities in 20 countries. Apart from universities, it is also used by high schools and publishers. They state that they are market leaders in Poland, Romania, and Ukraine. In 2018, they signed a Memorandum of Cooperation with the Ukrainian Ministry of Education and Science. The software searches in multiple databases and aggregators.

**Turnitin** was founded in 1999 by four students and grew to be an internationally known company. In 2014, they acquired the Dutch system Ephorus and "*joined forces*" (Ephorus, 2015). In 2019 they themselves were taken over by a US investment company, Advance (Turnitin, 2019). With a focus on institutional users only, they are used by 15,000 institutions in 150 countries. Turnitin uses its own crawler to search the web including also an archive of all previously indexed web pages (Turnitin, n.d.). Turnitin further compares the texts against published academic articles, as well as their own database of all assignments which have ever been submitted to the system, and optionally institutional databases. They are also developing many additional software tools for educators to use in teaching and giving feedback.

**Unicheck**, which declares itself to be a plagiarism checker, was launched in 2014 under the name Unplag. The brand name changed in 2017. It is operated by the company UKU Group ltd registered in Cyprus (Opencorporates, 2019). It is being used by 1,100 institutions in 69 countries, and apart from institutional users (high schools, higher education institutions, and business), they also offer their services for personal use. The pricing plans differ according to the type of user. Unicheck compares the documents with web content, open access sources and for business customers, also with their private library. They also claim to perform homoglyph (similar-looking character replacement) detection.

**Urkund,** which presents itself as a fully-automated text-recognition system for dealing with detection, handling, and prevention of plagiarism (Urkund, 2019), was founded in 1999. It is currently owned by a private equity fund, Procuritas Capital Investors VI, located in Stockholm. They claim to be a leader in the Nordic countries, and to have clients in 70 countries worldwide—mainly academic institutions and high schools, including over 800 of Sweden's high schools. They crawl the web "*with the aid of an ordinary search engine*" (Urkund, 2019) and they also compare the documents with student submissions to the system.

**Viper** presents itself as a plagiarism checker. It was founded in 2007. Viper focuses on all types of customers; the pricing is based on the pay-as-you-go principle. Currently, it is owned by All Answers Limited (2019), which according to the information at the website, gives an impression of an essay mill. It is interesting to see the progress in the way Viper uses the uploaded content on their "Terms and conditions" page. In 2016 the page stated *"[w]hen you scan a document, you agree that 9 months after completion of your scan, we will automatically upload your essay to our student essays database which will appear on one of our network of websites so that other students may use it to help them write their own essays"* (Viper, 2016). The time span was shortened to 3 months some time afterwards (Viper, 2019a). These paragraphs have been removed from the current version of the page (Viper, 2019b). On a different page, it is noted that *"[w]hen you scan your work for plagiarism using Viper Premium it will never be published on any of our study sites"* (Viper, 2019c). In e-mail communication, Viper claims that they are not using any essay without the author's explicit consent.

# 5. Coverage Results

This section discusses how much of the known text similarity was found by the systems. As they have various strengths and weaknesses, it is not possible to boil down the results to a single number that could easily be compared. Rather, the focus is on different aspects that will be discussed in detail. All tables in this section show the averages of the evaluation, therefore the maximum possible score is 5 and the minimum possible score is 0. Boldface indicates the maximum value achieved per each line, providing an answer to the question as to *which system performed best for this specific criterion*. All the values are shaded from red (worst) to dark green (best) with yellow being intermediate.

## 5.1 Language Comparison

Table 2 shows the aggregated results of the language comparisons based on the language sets. It can be seen that most of the systems performed better for English, Italian, Spanish, and German, whereas the results for Latvian, Slovak, Czech, and Turkish languages are poorer in general. The only system which found a Czech student thesis from 2010 which is publicly available from a university webpage, was StrikePlagiarism.com. The Slovak paper in an open-access journal was not found by any of the systems. Urkund was the only system that found an open-access book in Turkish. It is worth noting that a Turkish system, intihal.net, did not find this Turkish source.

Unfortunately, our testing set did not contain documents in Albanian or Slovenian, so we were not able to evaluate the potential strengths of the national systems (Akademia and DPV). And due to the restrictions on our account, it was not possible for us to process the Italian language in Akademia, although that should now be possible.

There are interesting differences between the systems depending on the language. PlagScan performed best on the English set, Urkund on Spanish, Slovak, and Turkish, PlagAware on German and StrikePlagiarism.com on Czech set. Three systems (PlagiarismCheck.org, PlagScan, and StrikePlagiarism.com) achieved the same maximum score for the Italian set.

**Table 2:** Coverage results according to language

| Language | Akademia | Copyscape | Docol©c | Dupli Checker | DPV | intihal.net | PlagAware | PlagiarismCheck.org | Plagiarism Software | PlagScan | StrikePlagiarism.com | Turnitin | Unicheck | Urkund | Viper |
|---|---|---|---|---|---|---|---|---|---|---|---|---|---|---|---|
| CZ | 1.4 | 1.5 | 1.1 | 0.9 | 0.0 | 0.7 | 1.8 | 1.5 | 1.5 | 1.5 | 2.2 | 1.6 | 1.0 | 1.9 | 1.0 |
| DE | 2.2 | 1.6 | 2.3 | 0.3 | 1.3 | 0.7 | **3.3** | 2.5 | 1.6 | 2.6 | 2.6 | 3.2 | 2.1 | 2.4 | 1.9 |
| EN | 2.9 | 3.3 | 2.8 | 0.9 | 0.9 | 1.3 | 3.2 | 2.1 | 1.2 | **4.0** | 3.0 | 3.7 | 3.2 | 3.3 | 1.3 |
| ES | 2.0 | 2.1 | 2.4 | 0.2 | 0.0 | 0.6 | 2.7 | 2.0 | 2.7 | 2.8 | 2.6 | 2.8 | 2.6 | **3.4** | 2.1 |
| IT | --- | 2.9 | 2.7 | 0.1 | 0.0 | 0.0 | 3.0 | **3.7** | 1.3 | **3.7** | **3.7** | 3.6 | 3.0 | 2.6 | 2.9 |
| LV | 1.5 | 2.3 | 1.8 | 0.6 | 0.0 | 0.0 | 2.3 | **3.8** | 1.6 | 2.7 | 2.6 | 1.2 | 2.3 | 3.2 | 2.3 |
| SK | 0.8 | 2.0 | 1.7 | 0.9 | 0.0 | 1.0 | 2.1 | 2.0 | 1.5 | 1.3 | 1.6 | 1.7 | 1.7 | **2.6** | 1.3 |
| TR | 2.0 | 1.5 | 1.3 | 0.3 | 0.5 | 1.0 | 2.1 | 1.7 | 1.1 | 1.9 | 1.8 | 2.0 | 1.4 | **3.1** | 1.2 |

Besides the individual languages, we also evaluated language groups according to a standard linguistic classification, that is, Germanic (English and German), Romanic (Italian and Spanish), and Slavic (Czech and Slovak). Table 3 shows the results for these language subgroups. Systems achieved better results with Germanic and Italic languages, their results are comparable. The results for Slavic languages are noticeably worse.

**Table 3:** Coverage results according to the language subgroups

| Language family | Akademia | Copyscape | Docol©c | Dupli Checker | DPV | intihal.net | PlagAware | PlagiarismCheck.org | Plagiarism Software | PlagScan | StrikePlagiarism.com | Turnitin | Unicheck | Urkund | Viper |
|---|---|---|---|---|---|---|---|---|---|---|---|---|---|---|---|
| Germanic | 2.5 | 2.5 | 2.5 | 0.6 | 1.1 | 1.0 | 3.3 | 2.3 | 1.3 | 3.3 | 2.8 | **3.5** | 2.6 | 2.9 | 1.6 |
| Romanic | --- | 2.5 | 2.6 | 0.2 | 0.0 | 0.3 | 2.8 | 2.9 | 2.0 | **3.3** | 3.2 | 3.2 | 2.8 | 3.0 | 2.5 |
| Slavic | 1.1 | 1.8 | 1.4 | 0.9 | 0.0 | 0.8 | 1.9 | 1.8 | 1.5 | 1.4 | 1.9 | 1.6 | 1.4 | **2.3** | 1.2 |

## 5.2 Types of Plagiarism Sources

This subsection discusses the differences between various types of sources with the results given in Table 4. The testing set contained Wikipedia extracts, open-access papers, student theses, and online documents such as blog posts. The systems generally yielded best results for Wikipedia sources. The scores between the systems vary due to their ability to detect paraphrased Wikipedia articles. Urkund scored the best for Wikipedia, Turnitin found the most open access papers, StrikePlagiarism.com scored by the best in the detection of student theses and PlagiarismCheck.org gave the best result for online articles.

Table 4: Coverage results according to the type of source

| Source | Akademia | Copyscape | Docol©c | Dupli Checker | DPV | intihal.net | PlagAware | PlagiarismCheck.org | Plagiarism Software | PlagScan | StrikePlagiarism.com | Turnitin | Unicheck | Urkund | Viper |
|---|---|---|---|---|---|---|---|---|---|---|---|---|---|---|---|
| Wikipedia | 2.3 | 2.3 | 2.2 | 0.9 | 0.8 | 0.9 | 2.4 | 2.8 | 1.5 | 2.4 | 2.6 | 2.6 | 2.3 | **3.0** | 2.0 |
| OA Paper | 1.4 | 1.1 | 1.3 | 0.4 | 0.4 | 0.5 | 1.6 | 1.1 | 1.2 | 1.5 | 1.2 | **1.8** | 1.5 | 1.7 | 0.8 |
| Student thesis | 0.2 | 0.1 | 0.1 | 0.1 | 0.1 | 1.0 | 0.3 | 0.1 | 1.3 | 0.3 | **2.1** | 0.1 | 0.1 | 0.8 | 0.1 |
| Online article | 0.6 | 2.2 | 1.8 | 0.7 | 0.5 | 0.6 | 1.8 | **2.8** | 1.4 | 2.5 | 2.3 | 1.6 | 2.2 | 2.0 | 2.0 |

Since it is assumed that Wikipedia is an important source for student papers, the Wikipedia results were examined in more detail. Table 5 summarizes the results from 3 x 8 single-source documents (one article per language) and Wikipedia scores from multi-source documents containing one-fifth of the text taken from the current version of Wikipedia. In general, most of the systems are able to find similarities to the text that has been copied and pasted from Wikipedia.

Ten years ago, Bretag & Mahmud (2009, p. 53) wrote that

> The text-matching facility in electronic plagiarism detection software is only suited to detect 'word-for-word' or 'direct' plagiarism and then only in electronic form. The more subtle forms of plagiarism, plus all types of plagiarism from paper-based sources, are not able to be detected at present.

Technological progress, especially in software development, advances rapidly. It is commonly expected that text-matching in the sense of finding both exact text matches and paraphrased ones should be a trivial task today. The testing results do not confirm this.

The results in Table 5 are quite surprising and indicate insufficient systems. The performance on plagiarism from Wikipedia disguised by a synonym replacement was generally poorer and almost no system was able to satisfyingly identify manual paraphrase plagiarism. This is surely due to both the immense number of potential sources and the exponential explosion of potential changes to a text.

Table 5: Coverage results for Wikipedia sources only

|  | Akademia | Copyscape | Docol©c | Dupli Checker | intihal.net | PlagAware | PlagiarismCheck.org | Plagiarism Software | PlagScan | DPV | StrikePlagiarism.com | Turnitin | Unicheck | Urkund | Viper |
|---|---|---|---|---|---|---|---|---|---|---|---|---|---|---|---|
| Copy-paste | 3.3 | 4.5 | 4.6 | 2.1 | 1.5 | 1.5 | 4.5 | 4.5 | 3.6 | 4.2 | 4.9 | 4.9 | 4.8 | **5.0** | 4.6 |
| Synonyms | 2.9 | 3.6 | 2.8 | 1.1 | 0.9 | 1.0 | 3.7 | 4.3 | 1.8 | 3.8 | 3.9 | 4.2 | 3.0 | **4.6** | 2.4 |
| Paraphrase | 2.0 | 1.6 | 1.2 | 0.1 | 0.5 | 0.7 | 1.9 | **2.8** | 0.8 | 2.1 | 1.9 | 2.0 | 1.3 | 2.7 | 1.0 |

## 5.3 Plagiarism Methods

The same aggregation as was done in Table 5 for Wikipedia was also done over all 16 single-source and eight multi-source documents. Not only copy & paste, synonym replacement and manual paraphrase were examined, but also translation plagiarism.

Translations were done from English to all languages, as well as from Slovak to Czech, from Czech to Slovak and from Russian to Latvian. The results are shown in Table 6, which confirms that software performs worse on synonym replacement and manual paraphrase plagiarism.

As has been shown in other investigations (Weber-Wulff et al., 2013) translation plagiarism is very seldom picked up by software systems. The worst performance of the systems in this test was indeed the translation plagiarism, with one notable exception—Akademia. This system is the only one that performs semantic analysis and allows users to choose the translation language. Unfortunately, their database—with respect to the languages of our testing—is much smaller than the database of other systems. However, the performance drop between copy-paste and translation plagiarism is much smaller for Akademia than for the other systems.

Given the very poor performance of the systems for translation plagiarism, it did not make sense to distinguish Google Translate and manual translation. The vast majority of the systems did not find text translated by either of them.

Some systems found a match in translation from Slovak to Czech, nevertheless, it was due to the words identical in both languages. For other languages, the examination of the system outputs for translation plagiarism revealed that the only match the systems found in translated documents was for the references. Matches in the references might be an indicator of translation plagiarism, but of course, if two papers use the same source, they should be written in an identical form if they follow the same style guide. This is an important result for educators.

Table 6: Coverage results by plagiarism method

| Method | Akademia | Copyscape | Docol©c | Dupli Checker | DPV | intihal.net | PlagAware | PlagiarismCheck.org | Plagiarism Software | PlagScan | StrikePlagiarism.com | Turnitin | Unicheck | Urkund | Viper |
|---|---|---|---|---|---|---|---|---|---|---|---|---|---|---|---|
| Copy-paste | 2.1 | 2.7 | 2.8 | 1.1 | 0.8 | 1.0 | 2.9 | 2.6 | 2.6 | 2.7 | 3.1 | 2.8 | 3.1 | **3.2** | 2.6 |
| Synonyms | 1.8 | 2.3 | 1.9 | 0.7 | 0.6 | 0.8 | 2.2 | 2.7 | 1.2 | 2.5 | 2.5 | 2.5 | 2.1 | **2.7** | 1.6 |
| Paraphrase | 1.3 | 1.2 | 1.0 | 0.4 | 0.5 | 0.7 | 1.4 | **1.7** | 0.8 | 1.5 | 1.3 | 1.5 | 1.1 | 1.6 | 0.9 |
| Translation | **1.1** | 0.0 | 0.1 | 0.0 | 0.2 | 0.3 | 0.2 | 0.2 | 0.3 | 0.0 | 0.4 | 0.3 | 0.2 | 0.5 | 0.3 |

## 5.4 Single-Source vs. Multi-Source Documents

One scenario that was considered in the test was when a text is compiled from short passages taken from multiple sources. This seems to be much closer to a real-world setting, in which plagiarism of a whole document is less likely, whereas 'patch-writing' or 'compilation' is a frequent strategy of student writers, especially second-language student writers (Howard, 1999, p. 117ff). Surprisingly, some systems performed differently for these two scenarios (see Table 7). To remove a bias caused by different types of sources, the Wikipedia-only portions were also examined in isolation (see Table 8), the results are consistent in both cases.

Table 7: Coverage results for single-source and multi-source documents

| All sources | Akademia | Copyscape | Docol©c | Dupli Checker | DPV | intihal.net | PlagAware | PlagiarismCheck.org | Plagiarism Software | PlagScan | StrikePlagiarism.com | Turnitin | Unicheck | Urkund | Viper |
|---|---|---|---|---|---|---|---|---|---|---|---|---|---|---|---|
| Single-source | 2.0 | 2.2 | 2.0 | 0.5 | 0.5 | 0.6 | 2.3 | 2.5 | 1.5 | 2.4 | 2.5 | 2.5 | 2.3 | **2.7** | 1.7 |
| Multi-source | 1.6 | 1.7 | 2.0 | 0.6 | 0.2 | 0.8 | **3.3** | 2.0 | 1.7 | 2.9 | 2.5 | 2.5 | 1.8 | 3.1 | 1.5 |

Table 8: Coverage results for Wikipedia in single-source and multi-source documents

| Wikipedia | Akademia | Copyscape | Docol©c | Dupli Checker | DPV | intihal.net | PlagAware | PlagiarismCheck.org | Plagiarism Software | PlagScan | StrikePlagiarism.com | Turnitin | Unicheck | Urkund | Viper |
|---|---|---|---|---|---|---|---|---|---|---|---|---|---|---|---|
| Single-source | 2.4 | 2.5 | 2.3 | 0.8 | 0.7 | 0.9 | 2.5 | 2.9 | 1.6 | 2.6 | 2.7 | 2.8 | 2.5 | **3.2** | 2.1 |
| Multi-source | 1.8 | 2.4 | 2.2 | 1.3 | 0.8 | 0.8 | **3.6** | 3.3 | 2.0 | 3.3 | 3.2 | 3.4 | 1.9 | **3.6** | 2.3 |

# 6. Usability

The usability of the systems was evaluated using 23 objective criteria which were divided into three groups of criteria related to the system workflow process, the presentation of the results, and additional aspects. The points were assigned based on researcher findings during a specific period of the time.

## 6.1 Workflow Process Usability

The first criteria group is related to the usability of the workflow process of using the systems. It was evaluated using the following questions:

1. Is it possible to upload and test multiple documents at the same time?
2. Does the system ask to fill in metadata for documents?
3. Does the system use original file names for the report?
4. Is there any word limit for the document testing?
5. Does the system display text in the chosen language only?
6. Can the system process large documents (for example, bachelor thesis)?

The results are summarized in Table 9. With respect to the workflow process, five systems were assigned the highest score in this category. The scores of only five systems were equal to or less than 3. Moreover, the most supported features are the processing of large documents (13 systems), as well as displaying text in the chosen language and having no word limits (12 systems). Uploading multiple documents is a less supported feature, which is unfortunate, as it is very important for educational institutions to be able to test several documents at the same time.

**Table 9**: Usability evaluation: Workflow process

| Workflow process | Akademia | Copyscape | Docol©c | Dupli Checker | DPV | intihal.net | PlagAware | PlagiarismCheck.org | Plagiarism Software | PlagScan | StrikePlagiarism.com | Turnitin | Unicheck | Urkund | Viper | Total |
|---|---|---|---|---|---|---|---|---|---|---|---|---|---|---|---|---|
| Upload multiple documents | 0 | 0 | 1 | 0 | 1 | 0 | 0.5 | 1 | 0 | 1 | 0 | 1 | 1 | 1 | 1 | 8.5 |
| No metadata required | 0 | 1 | 1 | 1 | 1 | 0 | 1 | 1 | 1 | 1 | 0 | 0 | 1 | 1 | 0 | 10 |
| Original filenames | 0 | 0 | 1 | 0 | 1 | 1 | 1 | 0 | 0 | 1 | 1 | 1 | 1 | 1 | 1 | 10 |
| No word limit | 1 | 0 | 1 | 0 | 1 | 1 | 1 | 1 | 0 | 1 | 1 | 1 | 1 | 1 | 1 | 12 |
| Text in a chosen language | 0 | 1 | 1 | 1 | 1 | 0 | 0 | 1 | 1 | 1 | 1 | 1 | 1 | 1 | 1 | 12 |
| Large document | 1 | 1 | 1 | 0 | 1 | 1 | 1 | 1 | 0 | 1 | 1 | 1 | 1 | 1 | 1 | 13 |
| Total | 2 | 3 | 6 | 2 | 6 | 3 | 4.5 | 5 | 2 | 6 | 4 | 5 | 6 | 6 | 5 | |

## 6.2 Result Presentation Usability

The presentation and understandability of the results reported by the systems were evaluated in a second usability criteria group. Since the systems cannot determine plagiarism, the results must be examined by one or more persons in order to determine if plagiarism is present and a sanction warranted. It must be necessary to download the result reports and to be able to locate them again in the system. Some systems rename the documents, assigning internal numbering to them, which makes it extremely difficult to find the report again. Many systems have different formats for online and downloadable reports. It would be useful for the report review if the system kept the original formatting and page numbers of the document being analyzed in order to ease the load of evaluation.

It is assumed that the vast majority of the universities require that the evidence of found similarities be documented in the report so that they can be printed out for a student's permanent record. This evidence is examined by other members of the committee, who may not have access to the system at a disciplinary hearing.

The results related to the presentation group are summarized in Table 10 and all criteria are listed below:

1. Reports are downloadable.
2. Results are saved in the user's account and can be reviewed afterwards.
3. Matched passages are highlighted in the online report.
4. Matched passages are highlighted in the downloaded report (offline).
5. Evidence of similarity is demonstrated side-by-side with the source in the online report.
6. Evidence of similarity is demonstrated side-by-side with the source in the downloaded report.
7. Document formatting is not changed in the report.
8. Document page numbers are shown in the report.
9. The report is not spoiled by false positives.

None of the systems was able to get the highest score in the usability group related to the test results. Two systems (PlagScan and Urkund) support almost all features, but six systems support half or fewer features. The most supported features are the possibility to download result reports and highlighting matched passages in the online report. Less supported features are a side-by-side demonstration of evidence in the downloaded report and in the online report, as well as keeping document formatting.

**Table 10**: Usability evaluation: Presentation of results

| Presentation of results | Akademia | Copyscape | Docol©c | Dupli Checker | DPV | intihal.net | PlagAware | PlagiarismCheck.org | Plagiarism Software | PlagScan | StrikePlagiarism.com | Turnitin | Unicheck | Urkund | Viper | Total |
|---|---|---|---|---|---|---|---|---|---|---|---|---|---|---|---|---|
| Downloadable report | 1 | 0 | 1 | 0 | 1 | 1 | 1 | 1 | 1 | 1 | 1 | 1 | 1 | 1 | 1 | 13 |
| Results saved | 1 | 0 | 1 | 0 | 1 | 1 | 1 | 1 | 0 | 1 | 1 | 1 | 1 | 1 | 1 | 12 |
| Highlights text match online | 1 | 0 | 0 | 1 | 1 | 1 | 1 | 1 | 1 | 1 | 1 | 1 | 1 | 1 | 1 | 13 |
| Highlights text match offline | 0 | 0 | 1 | 0 | 1 | 0 | 1 | 1 | 1 | 1 | 1 | 1 | 1 | 1 | 1 | 11 |
| Side-by-side comparison online | 0 | 0 | 0 | 1 | 1 | 0 | 0.5 | 0 | 0 | 1 | 0 | 0 | 0 | 1 | 0 | 4.5 |
| Side-by-side comparison offline | 0 | 0 | 0 | 0 | 0 | 0 | 0 | 0 | 0 | 0 | 0 | 0 | 0 | 1 | 0 | 1 |
| Document formatting not changed | 1 | 0 | 1 | 0 | 0 | 0 | 0 | 0 | 0 | 1 | 0 | 1 | 1 | 0 | 0 | 5 |
| Page numbers shown | 1 | 0 | 1 | 0 | 0 | 0 | 0 | 0 | 0 | 1 | 1 | 1 | 1 | 1 | 1 | 8 |
| No false positives | 0 | 1 | 1 | 1 | 1 | 0 | 1 | 1 | 0 | 1 | 1 | 1 | 0 | 1 | 1 | 11 |
| Total | 5 | 1 | 6 | 3 | 6 | 3 | 5.5 | 5 | 3 | 8 | 6 | 7 | 6 | 8 | 6 | |

## 6.3 Other Usability Aspects

Besides the workflow process and the result presentation, there are also other system usability aspects worth evaluating, specifically:
1. System costs are clearly stated in the system homepage.
2. Information about a free system trial version is advertised on the webpage.
3. The system can be integrated as an API to a learning management system.
4. The system can be integrated with the Moodle platform.
5. The system provides call support.
6. The call support is provided in English.
7. English is properly used on the website and reports.
8. There are no external advertisements.

In order to test the call support, telephone numbers were called from a German university telephone during normal European working hours (9:00–17:00 CET/GMT+1). A checklist (Weber-Wulff, 2019) was used to guide the conversation if anyone answered the phone. English was used as the language of communication, even when calling German companies. For intihal.net, the call was not answered, but an hour later the call was returned. StrikePlagiarism.org did not speak English, but organized someone who did. He refused, however, to give information in English and insisted that email support be used. Plagiarism Software publishes a number in Saudia Arabia, but returned the call from a Pakistani number. PlagAware only has an answering machine taking calls, but will respond to emails. The woman answering the Turnitin number kept repeating that all information was found on the web pages and insisted that since this was not a customer calling, that the sales department be contacted. Each of these systems was awarded half a point. Akademia published a wrong number on their web page as far as could be discerned, as the person answering the phone only spoke a foreign language. In case we did not reach anyone via phone and thus we could not assess their ability to speak English, we assigned 0 points for this criterion.

As shown in Table 11, only PlagiarismCheck.org and Unicheck fulfilled all criteria. Five systems were only able to support less than half of the defined features. The most supported features were no grammatical mistakes seen and no external advertisements. Problematic areas are not stating the system costs clearly, unclear possible integration with Moodle, and the lack of provision of call support in English.

Table 11: Usability evaluation: Other aspects

| | Akademia | Copyscape | Docol©c | Dupli Checker | DPV | intihal.net | PlagAware | PlagiarismCheck.org | Plagiarism Software | PlagScan | StrikePlagiarism.com | Turnitin | Unicheck | Urkund | Viper | Total |
|---|---|---|---|---|---|---|---|---|---|---|---|---|---|---|---|---|
| Costs clearly stated | 0 | 1 | 1 | 1 | 0 | 0 | 1 | 1 | 1 | 1 | 1 | 0 | 1 | 0 | 1 | 10 |
| Free trial advertised on web page | 0 | 1 | 0 | 1 | 0 | 0 | 1 | 1 | 1 | 1 | 1 | 0 | 1 | 0 | 0 | 8 |
| API integration | 0 | 1 | 1 | 0 | 0 | 0 | 1 | 1 | 0 | 1 | 1 | 1 | 1 | 1 | 0 | 9 |
| Moodle integration | 0 | 0 | 0 | 0 | 0 | 0 | 1 | 1 | 0 | 1 | 1 | 1 | 1 | 1 | 0 | 7 |
| Has call support | 0 | 0 | 1 | 0 | 0 | 0.5 | 0.5 | 1 | 0.5 | 1 | 0.5 | 0.5 | 1 | 1 | 0 | 7.5 |
| Support speaks English | 0 | 0 | 1 | 0 | 0 | 0.5 | 0 | 1 | 1 | 0 | 0 | 1 | 1 | 1 | 0 | 6.5 |
| No grammatical mistakes | 1 | 1 | 1 | 1 | 1 | 1 | 1 | 1 | 1 | 1 | 1 | 1 | 1 | 1 | 1 | 15 |
| No external advertisments | 1 | 1 | 1 | 0 | 1 | 1 | 1 | 1 | 1 | 1 | 1 | 1 | 1 | 1 | 1 | 14 |
| Total | 2 | 5 | 6 | 3 | 2 | 3 | 6.5 | 8 | 5.5 | 7 | 6.5 | 5.5 | 8 | 6 | 3 | |

# 7. Discussion

In the majority of the previous research on testing of text-matching tools, the main focus has been on coverage. The problem with most of these studies is that they approach coverage from only one perspective. They only aim at measuring the overall coverage performance of the detection tools, whereas the present study approaches coverage from four perspectives: language-based coverage, language subgroup-based coverage, source-based coverage, and disguising technique-based coverage. This study also includes a usability evaluation.

It must be noted that both the coverage and the usability scores are based on work that was done with potentially older versions of the systems. Many companies have responded to say that they now are able to deal with various issues. This is good, but we can only report on what we saw when we evaluated the systems. If any part of the evaluation was to be repeated, it would have to be repeated for all systems. It should be noted that similar responses have come from vendors for all of Weber-Wulff's tests, such as (Weber-Wulff et al., 2013).

It must be also noted that selection of usability criteria and their weights reflect personal experience of the project team. We are fully aware that different institutions may have different priorities. To mitigate this limitation, we have published all usability scores, allowing for calculations using individual weights.

## 7.1 Language-based Coverage

With respect only to the language-based coverage, the performance of the tools for eight languages was evaluated in order to determine which tools yield the best results for each particular language. The results showed that best-performing tools with respect only to coverage are (three systems tied for Italian):

- PlagAware for German;
- PlagScan for English, Italian;
- PlagiarismCheck.org for Italian, Latvian;
- Strikeplagiarism.com for Czech, Italian;
- Urkund for Slovak, Spanish, and Turkish.

It is worth noting that, in an overall sense, the text-matching tools tested yield better results for widely spoken languages. In the literature, language-based similarity detection mainly revolves around identifying plagiarism among documents in different languages. No study, to our knowledge, has been conducted specifically on the coverage of multiple languages. In this respect, these findings offer valuable insights to the readers. As for the language subgroups, the tested text-matching tools work best for Germanic languages and Romanic languages while results are not satisfactory for Slavic languages.

## 7.2 Source-based Coverage Testing

**Source-based coverage testing** was made using four types of sources; Wikipedia, open-access papers, a student thesis and online articles. For many students, Wikipedia is the starting point for research (Howard & Davies, 2009), and thus can be regarded as one of the primary sources for plagiarists. Since a Wikipedia database is freely available, it is expected that Wikipedia texts should easily be identifiable. Testing the tools with Wikipedia texts demonstrates the fundamental ability to catch text matches.

Three articles per language were created, each of which was made using a different disguising technique (copy & paste, synonym replacement and manual paraphrase) for all eight languages. The best performing tools for the sources tested over all languages were

- PlagiarismCheck.org for online articles,
- StrikePlagiarism.com for the student thesis (although this may be because the student thesis was in Czech),
- Turnitin for open-access papers,
- Urkund for Wikipedia

Since Wikipedia is assumed to be a widely used source, it was worth investigating Wikipedia texts deeper. The results revealed that the majority of tools are successful at detecting similarity with copy & paste from Wikipedia texts, except for Intihal.net, DPV and Dupli Checker respectively. However, a considerable drop was observed in synonym replacement texts in all systems, except for Urkund, PlagiarismCheck.org and Turnitin. Unlike other systems, Urkund, PlagiarismCheck.org and Turnitin yielded promising results in synonym replacement texts. This replicates the result of the study of Weber-Wulff et al. (2013), in which Urkund and Turnitin were found to have the best results among 16 tools.

As for the paraphrased texts, all systems fell short in catching similarity at a satisfactory level. PlagiarismCheck.org was the best performing tool in paraphrased texts compiled from Wikipedia. Overall, Urkund was the best performing tool at catching similarity in Wikipedia texts created by all three disguising techniques.

One aspect of Wikipedia sources that is not adequately addressed by the text-matching software systems is the proliferation of Wikipedia copies on the internet. As discussed in Weber-Wulff et al. (2013), this can lead to the appearance of many smallish text matches instead of one large one. In particular, this can happen if the copy of the ever-changing Wikipedia in the database of the software system is relatively old and the copies on the internet are from newer versions. A careless teacher may draw false conclusions if they focus only on the quantity of Wikipedia similarities in the report.

## 7.3 Disguising Technique-based Coverage

The next dimension of coverage testing is **disguising technique-based coverage**. In this phase, documents were created using copy & paste, synonym replacement, paraphrase, and translation techniques. In copy & paste documents, all systems achieved acceptable results except DPV, intihal.net and Dupli Checker. Urkund was the best tool at catching similarity in copy & paste texts. The success of some of the tools tested in catching similarity in copy & paste texts has also been validated by other studies such as Turnitin (Bull et al., 2001; Vani & Gupta, 2016; Kakkonen & Mozgovoy, 2010; Maurer et al., 2006) and Docol©c (Maurer et al., 2006).

For synonym replacement texts, the best-performing tools from copy & paste texts continued their success with a slight decline in scores, except for PlagiarismCheck.org which yielded better results in synonym replacement texts than copy & paste texts. Plagiarism Software and Viper showed the sharpest decline in their scores for synonym replacement. Urkund and PlagiarismCheck.org were the best tools in this category.

For paraphrased texts, none of the systems was able to provide satisfactory results. However, PlagiarismCheck.org, Urkund, PlagScan and Turnitin scored somewhat better than the other systems. PlagScan (Křížková, Tomášková, & Gavalec, 2016) and Turnitin (Bull et al., 2001) also scored well in paraphrased texts in some studies.

In translated texts, all the systems were unable to detect translation plagiarism, with the exception of Akademia. This system allows users an option to check for potential translation plagiarism. The systems detected translation plagiarism mainly in the references, not in the texts. This is similar to the previous research findings and has not been improved since then. For example, Turnitin and Docol©c have previously been shown not to be efficient in detecting translation plagiarism (Maurer et al., 2006). To increase the chances of detecting translation plagiarism, paying extra attention to the matches with the reference entries should be encouraged since matches from the same source can be a significant indicator of translation plagiarism. However, it should be noted that some systems may omit matches with the reference entries by default.

## 7.4 Multi-Source Coverage Testing

In the last phase of coverage testing, we tested the ability of systems to detect similarity in the documents that are compiled from multiple sources. It is assumed that plagiarised articles contain text taken from multiple sources (Sorokina, Gehrke, Warner & Ginsparg, 2006). This type of plagiarism requires additional effort to identify. If a system is able to find all similarity in documents which are compiled from multiple sources, this is a significant indicator of its coverage performance.

The multi-source results show that Urkund, the best performing system in single-source documents, shares the top score with PlagAware in multi-source documents, while Dupli Checker, DPV and intihal.net yielded very unsatisfactory results. Surprisingly, only the performance of two systems (Akademia and Unicheck) demonstrated a sharp decline in multi-source documents whereas the performance of ten systems actually improved for multi-source documents. This shows that the systems perform better in catching short fragments in a multi-source text rather than the whole document taken from a single source.

As for the general testing, the results are highly consistent with the Wikipedia results which contributes the validity of the single-source and multi-source testing. Again, in single-source documents, Urkund obtained the highest score, while PlagAware is the best performing system in multi-source documents. Dupli Checker, DPV and intihal.net obtained the least scores in both categories. Most of the systems demonstrated better performance for multi-source documents than for single-source ones. This is most probably explained by the chances the systems had for having access to a source. If one source was missing in the tool's database, it had no chance to identify the text match. The use of multiple sources gave the tools multiple chances of identifying at least one of the sources. This points out quite clearly the issue of false negatives: even if a text-matching tool does not identify a source, the text can still be plagiarized.

## 7.5 Overall Coverage Performance

Based on the total coverage performance, calculated as an average of the scores for each testing document, we can divide the systems into four categories (sorted alphabetically within each category) based on their overall placement on a scale of 0 (worst) to 5 (best).

- Useful systems - the overall score in [3.75–5.0]:
  There were no systems in this category
- Partially useful systems - the overall score in [2.5–3.75]:
  PlagAware, PlagScan, StrikePlagiarism.com, Turnitin, Urkund
- Marginally useful systems - the overall score in [1.25–2.5]:
  Akademia, Copyscape, Docol©c, PlagiarismCheck.org, Plagiarism Software, Unicheck, Viper
- Unsuited for academic institutions - the overall score in [0–1.25):
  Dupli Checker, DPV, intihal.net

## 7.6 Usability

The second evaluation focus of the present study is on usability. The results can be interpreted in two ways, either in a system-based perspective or a feature-based one, since some users may prioritize a particular feature over others. For the system-based usability evaluation, Docol©c, DPV, PlagScan, Unicheck, and Urkund were able to meet all of the specified criteria. PlagiarismCheck.org, Turnitin, and Viper were missing only one criterion (PlagiarismCheck.org dropped the original file names and both Turnitin and Viper insisted on much metadata being filled in).

In the feature-based perspective, the ability to process large documents, no word limitations, and using only in the chosen language were the features most supported by the systems. Unfortunately, the uploading of multiple documents at the same time was the least supported feature. This is odd, because it is an essential feature for academic institutions.

A similar usability evaluation was conducted by Weber-Wulff et al. (2013). In this study, they created a 27-item usability checklist and evaluated the usability of 16 systems. Their checklist includes similar criteria of the present study such as storing reports, side-by-side views, or effective support service. The two studies have eight systems in common. In the study of Weber-Wulff et al. (2013), the top three systems were Turnitin, PlagAware, and StrikePlagiarism.com while in the present study Urkund, StrikePlagiarism, and Turnitin are the best scorers. Copyscape, Dupli Checker, and Docol©c were the worst scoring systems in both studies.

Another similar study (Bull et al., 2001) addressed the usability of five systems including Turnitin. For usability, the researchers set some criteria and evaluated the systems based on these criteria by assigning stars out of five. As a result of the evaluation, Turnitin was given five stars for the clarity of reports, five stars for user-friendliness, five stars for the layout of reports and four stars for easy-to-interpret criteria.

The similarity reports are the end products of the testing process and serve as crucial evidence for decision makers such as honour boards or disciplinary committees. Since affected students may decide to ask courts to evaluate the decision, it is necessary for there to be clear evidence, presented with the offending text and a potential source presented in a synoptic (side-by-side) style, and including metadata such as page numbers to ease verifiability. Thus, the similarity reports generated were the focus of the usability evaluation.

However, none of the systems managed to meet all of the stated criteria. PlagScan (no side-by-side layout in the offline report) and Urkund (did not keep the document formatting) scored seven out of eight points. They were closely followed by Turnitin and Unicheck which missed two criteria (no side-by-side layout in online or offline reports).

The features supported most were downloadable reports and some sort of highlighting of the text match in the online reports. Two systems, Dupli Checker and Copyscape, do not provide downloadable reports to the users. The side-by-side layout was the least supported feature.

While four systems offer side-by-side evidence in their online reports, only one system (Urkund) supports this feature in the offline report. It can be argued that the side-by-side layout is an effective way to make a contrastive analysis in deciding whether a text match can be considered plagiarism or not, but this feature is not supported by most of the systems.

Along with the uploading process and the understandability of reports, we also aimed to address certain features that would be useful in academia. Eight criteria were included in this area:

- clearly stated costs,
- the offer of a free trial,
- integration to an LMS (Learning Management System) via API,
- Moodle integration (as this is a very popular LMS),
- availability of support by telephone during normal European working hours (9-15),
- availability of support by telephone in English,
- proper English usage on the website and in the reports, and
- no advertisements for other products or companies.

The qualitative analysis in this area showed that only PlagiarismCheck.org and Unicheck were able to achieve a top score. PlagScan scored seven points out of eight and was followed by PlagAware (6.5 points), StrikePlagiarism.com (6.5 points), Docol©c and Urkund (6 points). Akademia (2 points), DPV (2 points), Dupli Checker (3 points), intihal.net (3 points) and Viper (3 points) did not obtain satisfactory results.

Proper English usage was the most supported feature in this category, followed by no external advertisements. The least supported feature was clearly stated system costs, only six systems fulfilled this criterion. While it is understandable that a company wants to be able to charge as much as they can get from a customer, it is in the interests of the customer to be able to compare the total cost of use per year up front before diving into extensive tests.

In order to calculate the overall usability score, the categories were ranked based on their impact on usability. In this respect, the interpretation of the reports was considered to have the most impact on usability, since similarity reports can be highly misleading (also noted by Razı, 2005) when they are not clear enough or have inadequate features. Thus, the scores from this category were weighted threefold. The workflow process criteria was weighted twofold and the other criteria was weighted by one. The maximum weighted score was thus 47. Based on these numbers, we classified the systems into three categories (the boundaries for these categories were 35, 23, and 11:

- Useful systems: Docol©c, PlagScan, Turnitin, Unicheck, Urkund;
- Partially useful systems: DPV, PlagAware, PlagiarismCheck.org, StrikePlagiarism.com, Viper;
- Marginally useful systems: Akademia, Dupli Checker, Copyscape, intihal.net, Plagiarism Software.
- Unsuited for academic institutions: -

Please note that these categories are quite subjective, as our evaluation criteria are subjective and the weightings as well. For other use cases, the criteria might be different.

## 7.7 Combined Coverage

If the results for coverage and usability are combined on a two-dimensional graph, Figure 1 emerges. In this section, the details of the coverage and usability are discussed.

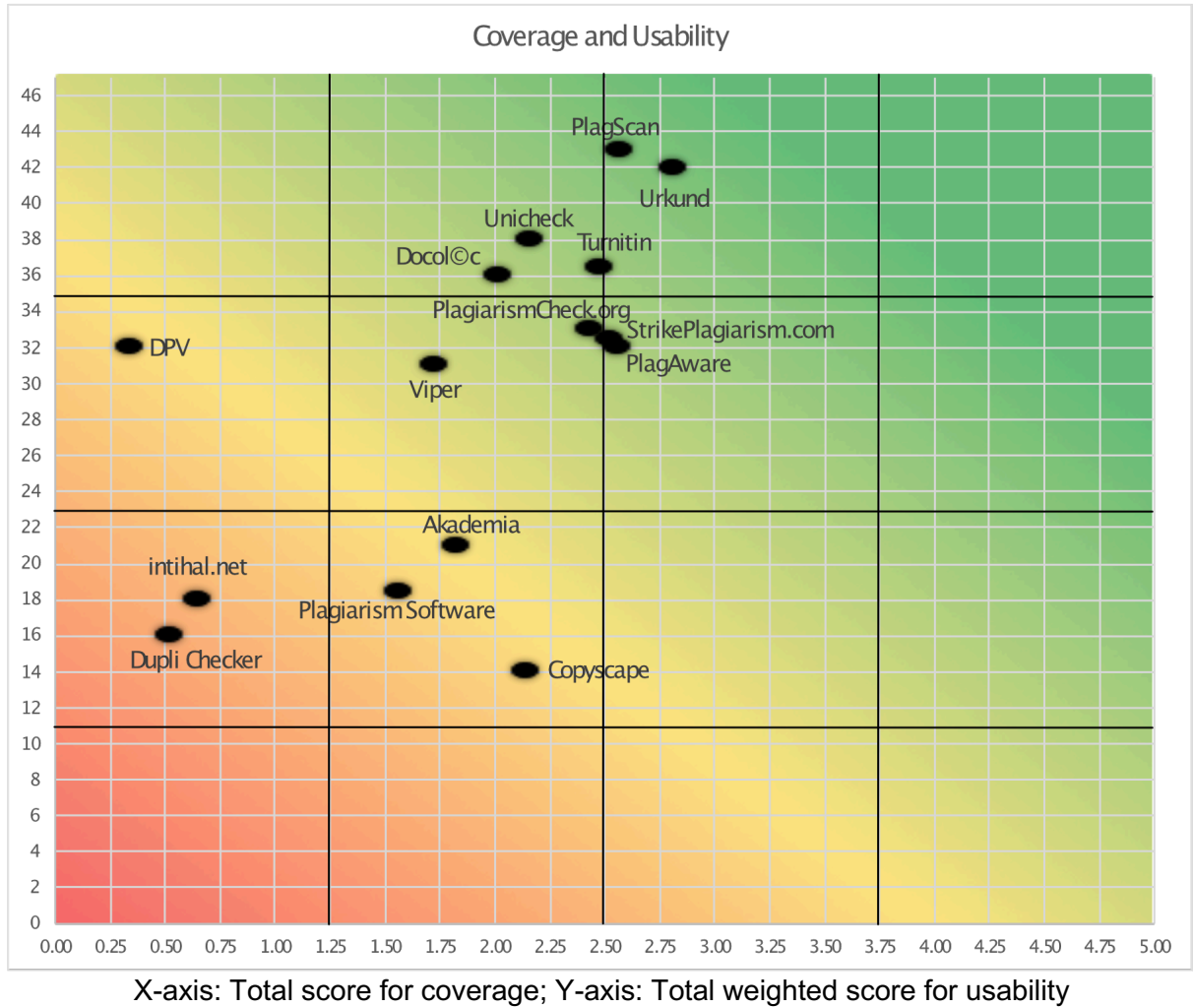

**Figure 1**: Coverage and Usability combined.

X-axis: Total score for coverage; Y-axis: Total weighted score for usability

Coverage is the primary limitation of a web-based text-matching tool (McKeever, 2006) and the usability of such a system has a decisive influence on the system users (Liu, Lo, & Wang, 2013). Therefore, Figure 1 presents a clear portrayal of the overall effectiveness of the systems. Having determined their criteria related to the coverage and usability of a web-based text-matching tool, clients can decide which system works best in their settings. Vendors are given an idea about the overall effectiveness of their systems among the tools tested. This diagram presents an initial blueprint for vendors to improve their systems and the direction of improvement.

One important result that can be seen in this diagram is that the usability performance of the systems is relatively better than their coverage performance (see Figure 1). As for coverage, the systems demonstrated at best only average performance. Thus, it has been shown that the systems tested fall short in meeting the coverage expectations. They are useful in the sense that they find some text similarity that can be considered plagiarism, but they do not find all such text similarity and they also suffer from false positives.

## 8. Conclusions and Recommendations

This study is the output of an intensive two-year collaboration and systematic effort of scholars from a number of different European countries. Despite the lack of external funding, the enthusiasm-driven team performed a comprehensive test of web-based text-matching tools with the aim to offer valuable insights to academia, policymakers, users, and vendors. Our results reflect the state of the art of text-matching tools between November 2018 and November 2019. Testing of the text-matching tools is not a new endeavour, however, previous studies generally have fallen short in providing satisfactory results. This study tries to overcome the problems and shortcomings of previous efforts. It compares 15 tools using two main criteria (coverage and usability), analyzing testing documents in eight languages, compiled from several sources, and using various disguising techniques.

A summary of the most important findings includes the following points:
- Some systems work better for a particular language or language family. Coverage of sources written in major languages (English, German, and Spanish) is in general much better than coverage of minor language sources (Czech or Slovak).
- The systems' performance varies according to the source of the plagiarized text. For instance, most systems are good at finding similarity to current Wikipedia texts, but not as good for open access papers, theses, or online articles.
- The performance of the systems is also different depending on the disguising technique used. The performance is only partially satisfactory in synonym replacement and quite unsatisfactory for paraphrased and translated texts. Considering that patchwriting, which includes synonym replacement and sentence re-arranging, is a common technique used by students, vendors should work to improve this shortcoming.
- The systems appear to be better at catching similarity in multi-source documents than single-source ones, although the test material was presented in blocks and not mixed on a sentence-by-sentence level.
- As for the usability perspective, this study clearly shows how important the similarity reports and how user-friendly the testing process of the systems are. The users can see which features are supported by the systems and which are not. Also, vendors can benchmark their features with other systems.

Based on our results, we offer the following recommendations for the improvement of the systems, although we realize that some of these are computationally impossible:

- Detect more types of plagiarism, particularly those coming from synonym replacement, translation, or paraphrase. Some semantic analysis results look promising, although their use will increase the amount of time needed for evaluating a system.
- Clearly identify the source location in the report, do not just report "Internet source" or "Wikipedia", but specify the exact URL and date stored so that an independent comparison can be done.
- Clearly identify the original sources of plagiarism when a text has been found similar to a number of different sources. For example, if Wikipedia and another page that has copied or used text from Wikipedia turn up as potential sources, the system should show both as possible sources of plagiarism, prioritizing showing Wikipedia first because it is more likely to be the real source of plagiarism. Once Wikipedia has been determined as a potential source, this particular article should be closely compared to see if there is more from this source.
- Avoid asking users to enter metadata (for example, author, title, and/or subject) in the system along with the text or file as mandatory information. It is good to have this feature available, but it should not be mandatory.
- Lose the single number that purports to identify the amount of similarity. It does not, and it is misused by institutions as a decision maker. Plagiarism is multi-dimensional and must be judged by an expert, not a machine. For example, a system could report the number of word match sequences found, the longest one, the average length of sequences, the number of apparent synonym substitutions, etc.
- Design useful reports and documentation. They must be readable and understandable both online and printed as a PDF. Special care should be taken with printed forms, as they will become part of a student's permanent record. It must show users the suspected text match side-by-side with the possible sources of plagiarism, highlighting the text that appears similar.
- Distinguish false positives from real plagiarism. Many of these false positives occur due to commonly used phrases within the context or language employed, or ignoring variant quotation styles (German or French quotation marks or indentation).

A number of important points for educators need to be emphasized:

1. Despite the systems being able to find a good bit of text overlap, they **do not determine plagiarism**. There is a prevalent misconception about these tools. In the literature, most of the studies use the term 'plagiarism detection tools'. However, plagiarism and similarity are very different concepts. What these tools promise is to find overlapping texts in the document examined. Overlapping texts do not always indicate plagiarism, thus the decision about whether plagiarism is present or not should never be taken on the basis of a similarity percentage. The similarity reports of these tools must be inspected by an experienced human being, such as a teacher or an academic, because all systems suffer from false positives (correctly quoted material counted as similarity) and false negatives (potential sources were not found).
2. Translation plagiarism can sometimes be found by a number of matches in references.
3. Another problem related to these tools is the risk of their possible cooperation with essay mills; this is because technically a company can store uploaded documents and share them with third parties. In the 'Terms and Conditions' sections of some tools, this notion is clearly stated. Uploading documents to such websites can cause a violation of ethics and laws, and teachers may end up with legal consequences. Thus, users must be skeptical about the credibility of the tools before uploading any documents to retrieve a similarity report.
4. It is necessary to obtain the legal consent of students before uploading their work to third parties. Since this legal situation can be different from country to country or even from university to university, make sure that the relevant norms are being respected before using such systems.
5. Because of European data privacy laws, for higher education institutions in the EU it must be certain that the companies are only using servers in the EU if they are storing material.
6. Teachers must make sure that they do not violate a non-disclosure agreement by uploading student work to the text-matching software.
7. Detecting plagiarism happens far too late in the writing process. It is necessary to institute institution-wide efforts to prevent academic misconduct and to develop a culture of excellence and academic integrity. This encourages genuine learning and shows how academic communication can be done right, instead of focusing on policing and sanctioning.

Considering both the number of participating systems and the number of testing documents and language variety, this paper describes the largest testing which has ever been conducted. We hope the results will be useful both for educators and for policymakers who decide which system to use at their institution. We plan to repeat the test in three years to see if any improvements can be seen.

# Appendix

These are the main contact URLs for the 15 systems evaluated.

| Tool | URL |
| --- | --- |
| **Akademia** | paneli.akademia.al |
| **Copyscape** | www.copyscape.com |
| **Docol©c** | www.docoloc.de |
| **DPV** | dpv.openscience.si |
| **Dupli Checker** | duplichecker.com |
| **intihal.net** | intihal.net |
| **PlagAware** | plagaware.com |
| **Plagiarism Software** | www.plagiarismsoftware.net |
| **PlagiarismCheck.org** | plagiarismcheck.org |
| **PlagScan** | www.plagscan.com |
| **StrikePlagiarism.com** | panel.strikeplagiarism.com |
| **Turnitin** | www.turnitin.com |
| **Unicheck** | unicheck.com |
| **Urkund** | www.urkund.com |
| **Viper** | www.scanmyessay.com |